\documentclass[10pt,twocolumn]{article}
\usepackage[top=1.8cm, bottom=1.8cm, left=1.8cm, right=1.8cm]{geometry}
\usepackage{times}  %
\usepackage[sort&compress,numbers]{natbib}  %
\usepackage[hyphens]{url}
\usepackage{graphicx}  %
\usepackage[keeplastbox]{flushend}
\usepackage[hyphens]{url}  %
\usepackage{graphicx} %
\usepackage{tikzsymbols}

\newcommand{\descr}[1]{\smallskip\noindent\textbf{#1}}

\usepackage{sectsty}
\sectionfont{\bfseries\Large\raggedright}

\usepackage{url}                                %
\usepackage{array,multirow,graphicx,adjustbox}  %
\usepackage{booktabs}                           %
\usepackage[utf8]{inputenc}
\usepackage[ruled,algosection,noend,linesnumbered]{algorithm2e}
\usepackage{float}
\usepackage{paralist}
\usepackage{amsmath}
\usepackage{amsfonts}
\usepackage{xcolor}
\usepackage{csquotes}
\usepackage{tabularx}
\usepackage{makecell}
\usepackage{pbox}
\usepackage[hang,flushmargin]{footmisc}
\usepackage{flushend}
\usepackage{xspace}
\usepackage{multicol, lipsum}
\usepackage[T1]{fontenc}

\usepackage{xcolor}
\usepackage{booktabs}
\usepackage{graphicx}
\usepackage{paralist}
\usepackage[small,labelfont=bf]{caption}
\usepackage{subfigure}

\let\oldbibliography\thebibliography
\renewcommand{\thebibliography}[1]{%
  \oldbibliography{#1}%
  \setlength{\itemsep}{2pt}%
}

\newcommand{\dspol}{/pol/\xspace}

\usepackage[hang,flushmargin]{footmisc}
\renewcommand{\footnotesize}{\fontsize{8}{9}\selectfont}
\usepackage[compact]{titlesec}
\titlespacing*{\section}{0pt}{*2.5}{2.5pt}
\titlespacing*{\subsection}{0pt}{*2}{2pt}
\usepackage{xspace}

\makeatletter
\def\url@leostyle{%
  \@ifundefined{selectfont}{\def\UrlFont{}}%
  {\def\UrlFont{}}%
}
\makeatother
\usepackage[hyphenbreaks]{breakurl}

\usepackage[bookmarks=true, bookmarksnumbered=true, colorlinks=true, linkcolor=linkcol, citecolor=citecol, urlcolor=urlcol, hypertexnames=true]{hyperref}

\definecolor{darkgreen}{RGB}{0, 100, 0}
\definecolor{linkcol}{rgb}{0.3,0,0}
\definecolor{citecol}{rgb}{0.3,0,0}
\definecolor{urlcol}{rgb}{0.3,0,0}

\makeatletter
\def\url@leostyle{%
  \@ifundefined{selectfont}{\def\UrlFont{\small}}%
  {\def\UrlFont{}}%
}
\makeatother
\begin{document}

\title{\bf A Multi-Platform Analysis of Political News Discussion \\and Sharing on Web Communities}

\author{Yuping Wang$^1$, Savvas Zannettou$^2$, Jeremy Blackburn$^3$, Barry Bradlyn$^4$,\\ Emiliano De Cristofaro$^5$, and Gianluca Stringhini$^1$\\[0.75ex]
{\normalsize $^1$Boston University, $^2$Max Planck Institute for Informatics, $^3$Binghamton University,}\\
{\normalsize $^4$University of Illinois at Urbana-Champaign, $^5$University College London}\\
{\normalsize -- iDRAMA Lab, https://idrama.science --}}

\date{}

\maketitle

\begin{abstract}
The news ecosystem has become increasingly complex, encompassing a wide range of sources with varying levels of trustworthiness, and with public commentary giving different spins to the same stories.
In this paper, we present a multi-platform measurement of this ecosystem.
We compile a list of 1,073 news websites and extract posts from four Web communities (Twitter, Reddit, 4chan, and Gab) that contain URLs from these sources.
This yields a dataset of 38M posts containing 15M news URLs, spanning almost three years.

We study the data along several axes, assessing the trustworthiness of shared news, designing a method to group news articles into \emph{stories}, analyzing these stories are discussed, and measuring the influence various Web communities have in that.
Our analysis shows that different communities discuss different types of news, with polarized communities like Gab and  /r/The\_Donald  subreddit disproportionately referencing untrustworthy sources.
We also find that fringe communities often have a disproportionate influence on other platforms w.r.t.~pushing narratives around certain news, for example about political elections, immigration, or foreign policy.
\end{abstract}

\section{Introduction}

The Web has facilitated the growth of fast-paced, online-first news sources.
It has also allowed users to actively contribute to and shape the discussion around news.
This %
creates an environment where journalists are not necessarily the arbiters of how a news story develops and spreads.
In today's ``hybrid'' media system~\cite{chadwick2011hybrid}, the popularity of a news story is also influenced by how users discuss it. %
Although such discussions usually happen organically, various actors from polarized online communities or state-sponsored troll farms might also attempt to manipulate them, e.g., by pushing~\cite{ferrara2017disinformation} or weaponizing~\cite{zannettou2017web} certain narratives.

Previous work studying the news ecosystem on the Web has mostly focused on Twitter, looking at single news articles~\cite{ratkiewicz2011truthy,shao2018spread,zhao2011comparing}, %
or on the discussion surrounding a small set of events~\cite{conover2011political,starbird2017examining,wilson2018assembling}. %
Moreover, efforts to study the intertwined relationship between news coverage and social media discussions have been limited to direct quotes from news articles posted on Twitter~\cite{leskovec2009memetracking}, %
or on how Web communities influence each other in spreading {\em single} news URLs~\cite{zannettou2017web}.
Overall, despite the crucial role played by online news in our society, we still lack computational tools to monitor how news stories unfold and are discussed across multiple online services.
In this paper, we %
present a longitudinal study of how news is disseminated across \emph{multiple} Web communities.
We introduce an analysis pipeline (which is independent of the  data sources and thus reusable), consisting of different components to: 1) collect data, 2) extract named entities, 3) group articles belonging to the same story, and 4) estimate influence of a Web community on other ones.

We instantiate the pipeline by focusing on a mix of mainstream and fringe communities -- namely, Twitter, Reddit, 4chan's Politically Incorrect board (\dspol), and Gab -- and extract 38M posts including 15M news URLs, spanning a period of almost three years.
We use named entity extraction to analyze what types of news these communities discuss.
We also study the interplay between news discussion and the trustworthiness of the news sources cited using NewsGuard~\cite{newsguardtech}, a trustworthiness assessment site compiled by professional fact checkers.

Next, we perform community detection to group together related articles into news stories, and study how they are discussed on different Web communities.
To this end, we use GDELT, a dataset that labels global news events~\cite{leetaru2013gdelt}.
Because of GDELT's focus on politics, our measurement also concentrates on political news stories.
Finally, to study the influence that different Web communities have on each other in spreading news stories, we use Hawkes Processes~\cite{hawkes1971spectra}.
These allow us to estimate which news stories are organically discussed on the Web, and for which ones certain communities exercise a significant influence in spreading them.

\begin{figure*}[t]
\includegraphics[width=\textwidth]{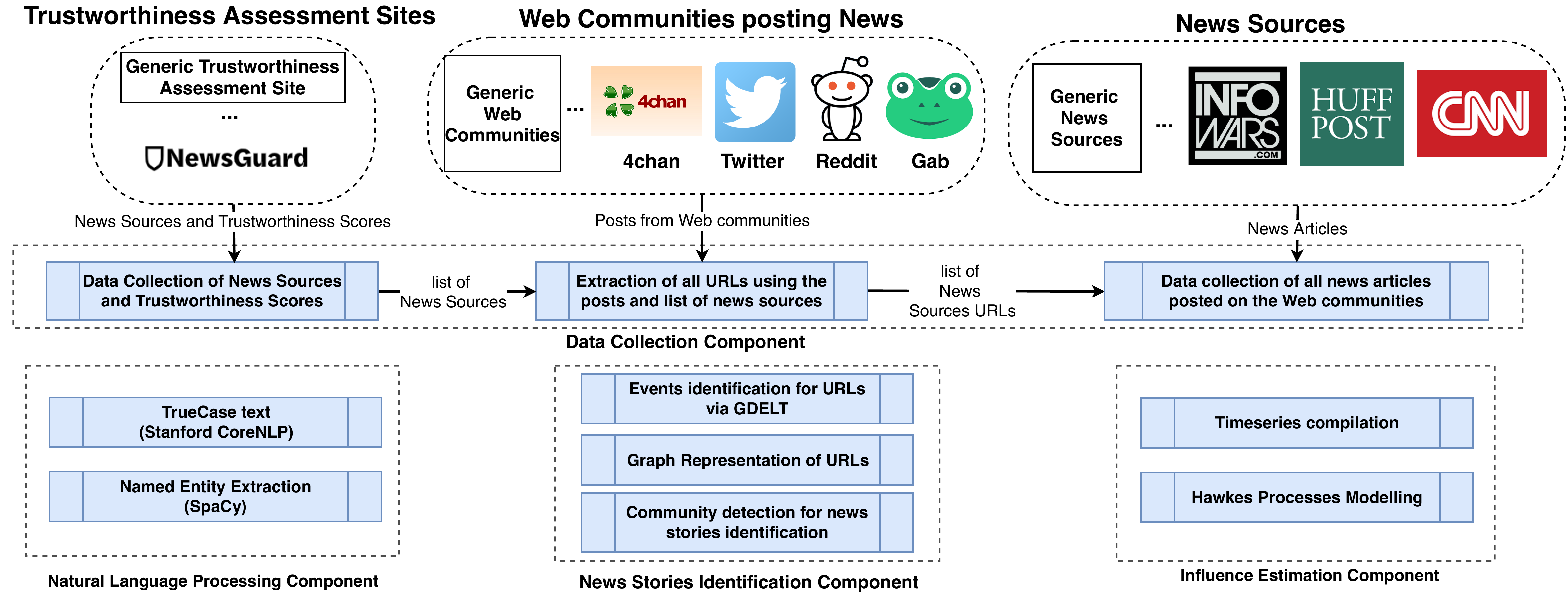}
\caption{High-level overview of our processing pipeline.}
\label{fig:pipeline}
\end{figure*}

Our analysis yields the following main findings:
\begin{compactitem}
\item When discussing news, different Web communities post URLs to news outlets with varying levels of trustworthiness. In particular, Gab and /r/The\_Donald (subreddit) prefer untrustworthy ones compared to Reddit, Twitter, and 4chan.
\item Some communities are particularly influential in the dissemination of news.
While large ones like Twitter and Reddit have a consistent influence on the others, relatively small/fringe communities like /r/The\_Donald have a much larger external influence, affecting the posting of stories at a rate that is much larger than their relative size would suggest.
\item Some topics are very popular across the board, however, different communities focus on different narratives about the same story.
For instance, /r/The\_Donald and \dspol are particularly influential in spreading anti-immigration rhetoric and conspiracy theories.
\end{compactitem}

\section{Methodology \& Datasets}\label{sec:methodology}
In this section, we present the methodology used to analyze the appearance and discussion of news articles across multiple Web communities, as well as the datasets we collect in the process.
\figurename~\ref{fig:pipeline} presents a high-level overview of our pipeline, which consists of four components:\\

\begin{compactenum}
\item {\em Data Collection}: selecting news sources, collecting related posts from social media, and gathering the news content.

\item {\em Natural Language Processing}: identifying the ``entities'' mentioned in news articles and discussions in Web communities. %

\item {\em News Stories Identification}: grouping articles belonging to the same story. %

\item {\em Influence Estimation}: assessing the influence of each Web community on others with respect to news stories. %
\end{compactenum}

\subsection{Data Collection}\label{sec:data_collection}
This component is used to select a set of suitable news sources, determine their trustworthiness, and retrieve posts on different social networks including URLs to these sources.
While our methodology is general and can be used with any data source,  in the following, we describe it along with the specific services selected for this work.

\descr{News Sources.}
\label{sec:news_dataset}
Previous work has mostly relied on pre-determined lists of websites~\cite{zannettou2017web, grinberg2019fake, budak2019happened, allcott2019trends}.
However, this incurs important limitations, as the popularity of news sites varies over time~\cite{scheitle2018long}, and low-reputation sites are often ephemeral. %
For instance, out of the 54 ``alternative'' news sites studied in~\cite{zannettou2017web},
only 23 are still active as of October 2020. %
Thus, we opt to take a more systematic approach.
First, we gather popular domains using the top 30K websites from the Majestic list~\cite{majestic-million} as of February 2019.
Then, we use the VirusTotal API~\cite{virustotal}, a service that provides domain categorization, to select only the domains categorized as \emph{news and media} or \emph{news}.
Note that VirusTotal's categorization is not exempt from mistakes; e.g., domains like \url{ananova.com}, \url{adbusters.org}, and \url{cagle.com} are misclassified as news sites.
To further refine this list, we use the NewsGuard API~\cite{newsguardtech}, a service that ranks news sources based on their trustworthiness, and restrict our analysis to news sites that are rated by NewsGuard as of February 2019, i.e., before it became a paid service.
As a result, we obtain a total of 1,073 news websites.

\descr{Source Trustworthiness.}
We use NewsGuard~\cite{newsguardtech} also to characterize the trustworthiness of a news Website, as it provides credibility/transparency scores. %
The scores are based on nine journalistic criteria, focused on different aspects (e.g., whether a news site consistently publishes false content, uses deceptive headlines, etc.) that do not take into account political leanings, and range from 0 to 100.
If the score is no less than 60, the news source is labeled as trustworthy, and untrustworthy otherwise~\cite{rating-process-criteria}.
NewsGuard's evaluation is conducted by a team of experts~\cite{rating-process-criteria}, and this manual vetting provides us with reasonable confidence in its accuracy. %
The threshold of 60 is pre-defined by NewsGuard, and its assessment evaluation is designed to fix the threshold first and then assign the points for each criterion according to this threshold~\cite{rating-process-criteria}.
Therefore, we stick to this threshold.

Overall, we are confident in the meaningfulness of NewsGuard scores, as a recent Gallup study~\cite{Gallup-NewsGuards} has confirmed it as a useful tool to help readers identify untrustworthy news outlets.
Approximately 600 news outlets in the US and Europe have refined their editorial practices to get higher scores~\cite{newsguards-first-year}.
Also, NewsGuard has been working with researchers~\cite{resnick2018iffy, zhou2020recovery, norregaard2019nela}, libraries, Web browsers, and service providers to increase the transparency of news credibility assessments~\cite{newsguards-first-year}.
News outlets are evaluated in a transparent way, as detailed information is published in the corresponding ``nutrition label'' page where readers can find the reasons for the judgment~\cite{sample-nutrition-labels}.
{\em NB: The full list of news sources used in this paper along with their NewsGuard scores is available, anonymously, from~\cite{open}.}

\descr{Web Communities.}\label{sec:communities}
We retrieve social media posts that include URLs from the 1,073 news sources in our dataset.
Our selection is based on highlighting the interplay and the influence between different online communities
instead of political leanings.
While our pipeline can include any platforms, in this paper, we focus on a few Web communities: Twitter, Reddit, 4chan, and Gab. %
That is, we study both mainstream communities like Twitter as well as ``fringe'' ones like 4chan.
In particular, we turn to 4chan's Politically Incorrect board (\dspol) as prior work shows it is an influential actor with respect to the dissemination of news~\cite{zannettou2017web} and memes~\cite{zannettou2018origins}.
We also include Gab because it is an emerging community marketed as a ``free speech haven,'' which is heavily used by alt- and far-right users~\cite{zannettou2018gab}.
As for Reddit, we also choose to study /r/The\_Donald as a separate community, since previous research has highlighted its  influence in spreading information on the Web~\cite{flores2018mobilizing,zannettou2017web}.\footnote{Note that /r/The\_Donald was banned by Reddit in 2020~\cite{ban_td}. However, our research was done before this ban.}

Table~\ref{tbl:datasets} provides a summary of the number of posts and unique news URLs for each community.
Next, we describe the data that we collect for each Web community in detail.

\begin{table}[t!]
\centering
\vspace{0.2cm}
\resizebox{\columnwidth}{!}{
\begin{tabular}{@{}lrrrr@{}}
\toprule
 & \multicolumn{2}{c}{\textbf{\#Posts}} & \multicolumn{2}{c}{\textbf{\#Unique URLs }}\\
\textbf{Community}   & \multicolumn{1}{l}{\textbf{Trust.}} & \multicolumn{1}{l}{\textbf{Untrust.}} & \multicolumn{1}{l}{\textbf{Trust.}} & \multicolumn{1}{l}{\textbf{Untrust.}}\\ \midrule
       Twitter &                  7,123,715 &                    686,497 &                     3,893,357 &                          291,354 \\
        Reddit &                 23,605,406 &                   1,342,429 &                    11,170,005 &                          612,213 \\
    /r/The\_Donald &                   528,142 &                    190,742 &                      385,384 &                          122,204 \\
         4chan &                   458,431 &                     75,705 &                      275,422 &                           37,472 \\
           Gab &                  2,369,149 &                   2,265,336 &                      749,547 &                          385,317 \\ \midrule
{\em Total} & 33,556,092 & 4,369,923 & 14,636,451& 984,812\\
\bottomrule
\end{tabular}%
}
\caption{Overview of our datasets. For each community, we report the number of posts that include URLs to trustworthy and untrustworthy news sources.}
\label{tbl:datasets}
\end{table}

\descr{\em Twitter.} We collect tweets made available through the 1\% Streaming API between Jan 1, 2016 and Oct 31, 2018.
Note that, due to failure on our data collection infrastructure, we have some gaps in our dataset; specifically, 1) Dec 4-11, 2016; 2) Dec 25, 2016 to Jan 08, 2017; 3) Dec 17, 2017 to Jan 28, 2018; and 4) Sep 20 to Oct 31, 2018.
We extract all tweets containing URLs to one of the news sources we study, collecting 7M tweets containing URLs to trustworthy news sources and 686K tweets containing URLs to untrustworthy news sources.
The total number of unique (news) URLs is 3.9M and 291K for, respectively, trustworthy and untrustworthy news.

\descr{\em Reddit and /r/The\_Donald.}
For Reddit, we use the monthly dumps available from Pushshift~\cite{baumgartner2020pushshift}.
We collect all submissions and comments from Jan 1, 2016 to Oct 31, 2018, and extract all submissions and comments that include a URL to the news sources in our data.
For the whole Reddit dataset, we find 23M and 1.3M posts with trustworthy and untrustworthy news URLs, respectively, while for /r/The\_Donald, 528K posts with trustworthy  and 190K with untrustworthy news URLs.
The number of unique URLs is 11.2M and 612K, respectively, for trustworthy and untrustworthy news for the entirety of Reddit and 385K and 122K for /r/The\_Donald.
Note that the Reddit dataset also includes /r/The\_Donald, as we aim to study the dynamics of Reddit as a whole. Nevertheless, even though /r/The\_Donald is a very small subset, and as such it has negligible effect on the analysis, we remove it from Reddit for our influence estimation presented in Section~\ref{sec:influence_results} to guarantee the quality of the results of the underlying statistical model.

\descr{\em 4chan's \dspol.} We obtain all posts on 4chan's Politically Incorrect board (\dspol) between Jun 30, 2016 and Oct 31, 2018 from~\cite{papasavva2020raiders}.
We extract all posts containing URLs to one of the news sources, collecting 458K and 75K posts with URLs to trustworthy and untrustworthy news sources, respectively, for a total of 275K and 37K, respectively, unique URLs.

\descr{\em Gab.} We use the data collection methodology presented in~\cite{zannettou2018gab} to collect Gab posts from Aug 10, 2016, and Oct 31, 2018.
Once again, we extract posts that include URLs to trustworthy and untrustworthy news sources, collecting 2.3M posts containing trustworthy news URLs and 2.2M posts containing untrustworthy news URLs.

\smallskip In total, we extract 15.6M URLs, 14.6M pointing to trustworthy and 984K to untrustworthy news sources, posted on the five Web communities.
Note that the Twitter and Reddit datasets start a few months earlier (January 2016) than Gab and 4chan.
This is because the authors of~\cite{papasavva2020raiders} began collecting 4chan data in June 2016, and 4chan data is ephemeral, therefore it is not possible to retrieve older posts.
Gab, on the other hand, was launched in August 2016.

\descr{News Content.}
\label{sec:news articles dataset}
Next, we collect the {\em content} of the 15.6M news articles %
using the Newspaper library for Python3~\cite{newspaper3k}, which, given a URL, retrieves the text from an article.
For sanity check, we have one author randomly select 20 URLs from the Web communities dataset, download the text of these articles, and then manually compare the text with the content on the Web page.
For 18 articles, the library downloads the full content, while, for 2, the text is partially downloaded---specifically, one misses the first two paragraphs, and the other only has the first paragraph.
This provides us with reasonable confidence of the effectiveness of Newspaper to download the text of online news articles.
Although the text of a small number of articles might be downloaded partially, this has a limited effect on our analysis since this text is only used when performing named entity recognition.
Since the first paragraph of an article usually provides the most important information about the covered story followed by details, referred to as ``inverted pyramid''style~\cite{inverted_pyramid_journalism}, we expect the overall effect of this issue to be small. See Section \ref{sec:general_communities} for more details.

Note that we are unable to retrieve the content of about 1.4M articles due to server-side problems (e.g., the article was deleted from the server or the server was down at that time) and about 1M articles because of paywalls.
For the latter, manual inspection shows that paywalls typically trigger a set of standard sentences being displayed instead of the actual news content (e.g., ``sign up for a new account and purchase a subscription'').
Thus, we parse results to exclude articles containing these sentences.
At the end, we gather the text of 13M articles, 12M from trustworthy sources and almost 1M from untrustworthy ones.

\descr{Ethical Considerations.} Our datasets only include publicly available data and we do not interact with users.
As such, this is not considered human subjects research by our IRB.
Also, we follow standard ethical guidelines~\cite{rivers2014ethical}, encrypt data at rest, and make no attempt to de-anonymize users.

\subsection{Natural Language Processing}
\label{sec:natural}

We now describe the NLP component, which we use to extract meaningful named entities that are referenced both on news articles and on discussions on several Web communities.
Our NLP component involves two models: 1)~a true case model that predicts and converts text into its correct case (e.g., ``donald trump is the president'' is converted to ``Donald Trump is the president''); and
2)~a named entity detection model that extracts known named entities from text along with an associated category (i.e., whether the extracted entity is a person, an organization, etc.).
The former is necessary since the latter is case sensitive.

\descr{True Case Model.}
We use \emph{TrueCaseAnnotator} from the Stanford CoreNLP toolkit~\cite{manning2014stanford}.
This converts the case of the text to match as it should appear in a well-edited format (e.g., ``united states'' becomes ``United States'' ), using a discriminative model built on Conditional Random Fields (CRFs)~\cite{lafferty2001conditional}.

\descr{Named Entity Extraction Model.}
To obtain named entities, we rely on the SpaCy library~\cite{spacy.io} and the \emph{en\_core\_web\_lg} model.
We choose this model since it is %
trained on the largest available dataset.
The named entity detection model leverages millions of Web entries consisting of blogs, news articles, and comments to detect and extract a wide variety of entities from text, ranging from people to countries and events (see~\cite{named-entities} for a list of all the supported types of entities).
The model relies on Convolutional Neural Networks (CNNs), trained on the OntoNotes dataset~\cite{LDC2013T19}, as well as Glove vectors~\cite{pennington2014glove} trained on the Common Crawl dataset~\cite{commoncrawl}.

\subsection{News Stories Identification}
\label{sec:news stories identification component}

The news articles in our dataset cover various aspects ranging from politics to entertainment.
Among all categories, we focus on politics because previous work showed that these stories are often discussed differently on different online communities~\cite{zannettou2017web} and are often used to spread disinformation narratives~\cite{starbird2017examining,wilson2018assembling,zannettou2019let}.
Therefore, we design a news stories identification component to group political news articles covering the same ``story.''

We use the definition by Marcelino et al.~\cite{marcelino2019benchmark}, whereby every news ``story'' is composed of several story ``segments.''
In a nutshell, we perform three tasks: 1) we identify segments using the GDELT dataset~\cite{leetaru2013gdelt}; 2) we build a graph where news articles are nodes and edges are common segments discussed in them; and 3) we perform community detection on the graph to identify articles that discuss the same story.

\descr{Event Identification with GDELT.}
\label{sec:events identification}
In this study, we use events identified by GDELT~\cite{leetaru2013gdelt} in an article as a   news story ``segment.''
GDELT is
a dataset containing event information for articles (published between Oct 30, 2015 and Nov 3, 2018) covering political news stories.
GDELT's focus on politics makes it the ideal candidate for our analysis pipeline.
We find 31M unique news URLs belonging to the 1,073 domains that we study in the GDELT dataset, composed of 30M unique trustworthy news URLs and 712K unique untrustworthy news URLs.
For each URL, GDELT lists \emph{events} (e.g.,  ``Egyptian Minister of Foreign Affairs Mohamed Orabi attended the summit yesterday''~\cite{leetaru2013gdelt}), which are each assigned a globally unique identifier (``Event ID'').
The event extraction is performed at the sentence level by an automated coding engine called TABARI~\cite{leetaru2013gdelt}, which identifies the actors involved in the event, the action performed, and where the event happened.
The result of this is that two different sentences referring to the same event are given an identical event ID, and each sentence is an \emph{event mention} of this event ID~\cite{GDELT-EventCodebook}.
When identifying  an event from a sentence, GDELT also gives a confidence score to the event mention, ranging from 10 to 100$\%$ (in 10\% increments), representing how sure the system is that this sentence indeed corresponds to that event ID~\cite{GDELT-EventCodebook}. %

To extract the segments associated to the news articles in our dataset, we first look up which of the URLs in our dataset are present in GDELT after a number of pre-processing steps, such as expanding shortened URLs, removing the query string as well as the {\tt www} prefix or slash suffixes from the URLs.
Then, we extract the list of corresponding events for each matched URL. %
We find 3M URLs in the dataset, comprising 24.6M event mentions (i.e., story segments).
As we mentioned, the reason why GDELT focuses on political news, and therefore does not cover all news in our dataset, which are often about other topics like sports or entertainment.

\descr{Graph Representation of Stories.}
After labeling news URLs with events that are relevant to them, we build a graph linking single articles with common events they cover.
In other words, if two articles share one event, these two articles are ``related'' and
the more events two articles share, the closer their content is.
The graph is built as follows:
1)~We treat each URL (stripped of its parameters) as a node. %
2)~We remove all event mentions for which GDELT has a confidence lower than 60$\%$.
We select the $60\%$  threshold as a tradeoff between removing too many events and ensuring high precision in event identification.
As discussed later, event mentions with low confidence are not reliable and keeping them in the graph ends up producing poor results;
as a result we remove 18.2M out of the 24.6M event mentions.
3)~If two URLs share at least one event, we build an edge between the two nodes.
4)~The edge weight is computed as the number of unique event IDs that two URLs share.

\descr{Community Detection.}
Two URLs that share one or more events are not guaranteed to cover the same story.
To further refine the association between common events and news stories, we apply community detection on the graph.
We then consider URLs to belong to the same news story if they are part of the same community. %
We apply the Louvain Method~\cite{blondel2008fast}, which allows to efficiently find communities in sparse graphs like the one we are dealing with: our graph is composed of 3M nodes and  1.6M edges.
First, though, we prune edges with weight lower than a threshold $d$, %
since, upon manual inspection, we find that the GDELT events include some noise, possibly due to crawler or event extraction faults.
In the following, we discuss how we select the value of the threshold $d$ for our experiments.

\descr{Selecting the Story Edge Weight Threshold.}
\label{sec:story edge weight validation}
To select the threshold $d$,
we first discard URLs with more than 60 unique event IDs (220 URLs in total).
We find that these results are due to errors in GDELT's crawling process; when manually inspecting these URLs, we find that the content is mostly homepages of news outlets, including numerous headlines and therefore flagged with multiple spurious event IDs. %
Further, we remove communities whose URLs are from a single domain only,
since by manually looking at the clusters, we find URLs within such a cluster are published on the same day and share several events even though their texts are totally different.

Then, to select the threshold $d$, we perform the following steps with $d \in \{1,2,3,4\}$.
We apply the Louvain Method, obtain the corresponding communities of URLs, and randomly select 20 communities with size larger than 10.
These samples are independently inspected by two authors to determine if the articles in them belong to the same story.
After comparing their results, the two authors
agreed that all samples with both $d=3$ and $d=4$ have a precision (i.e., the number of correctly grouped articles vs. all articles in the community)
above $90\%$.
Since $d=3$ produces more communities than $d=4$ (43K vs 26K),
we settle on $d=3$. %

To further verify the appropriateness of our parameter choice, we also run an experiment in which we keep the threshold at $d=3$, but this time we do not prune events with confidence lower than 60$\%$, keeping them in the graph.
As a result of this experiment we obtain 115k communities in total, which is much higher than on the pruned graph (43K).
However, upon manual inspection, we observe that the quality of the identified communities is not satisfactory in this case.
(Upon checking a sample of 20 communities, only 13 of them had a precision higher than $90\%$).

\descr{Alternatives to GDELT.} Note that we also tested alternatives to GDELT as external ``ground truth.''
More specifically, we group articles based on the TF-IDF~\cite{manning2008introduction} of their text and DBSCAN clustering~\cite{ester1996density}. %
However, manual analysis reveals that the performance of these methods is substantially
worse. %
An alternative approach would have been to use topic modeling, e.g., LDA~\cite{blei2003latent}. %
However, these methods are most effective when modeling topics that are broader than fine-grained news stories, and are therefore less appropriate than our approach in this case.
The reason is that features from LDA, as well as TF-IDF, are obtained at the \emph{word} level.
So keywords shared by two different stories can interfere with the clustering result while features from our method are obtained from the \emph{sentence} level (i.e., events in the stories), which avoids this issue.
One example is a  pair of stories found in our result:
``Donald Trump's call to punish flag burners caters to voters in his base'' and ``air conditioning company Carrier says it has deal with Trump to keep jobs in Indiana.''
Both stories are from Nov 29, 2016 (and thus cannot be distinguished by  date), and their text share some key word candidates: donald, trump, president-elect, tweet, which makes it difficult for LDA and TF-IDF to distinguish between them.

\subsection{Influence Estimation}\label{sec:temporal}
After grouping articles into news stories, we are interested in studying the temporal characteristics of how these stories are discussed in various communities of interest.
More specifically, we aim to understand and measure the interplay between multiple Web communities with respect to the news stories they share.
To do so, %
we create a timeseries that captures the cascades of each news story per Web community.
After obtaining the timeseries, we model the interplay between Web communities using a statistical framework known as Hawkes Processes~\cite{hawkes1971spectra}, which lets us quantify the influence that each Web community has on the others with respect to the dissemination of news stories. %

\descr{Timeseries compilation.}
As a first step, we organize our data into timeseries.
For each community of interest, we focus on the news stories that appear at least 100 times in our datasets (0.84\% of all stories).
This restriction helps to ensure the quality of the data under analysis.
Next, for each news story $i$, on each Web community $k$, we build a timeseries $u_{ik}(t)$, whose value is the number of occurrences of news URL related to a specific news story per $t$ hours.

\descr{Hawkes Processes} are self-exciting temporal point processes~\cite{hawkes1971spectra} that describe how events (in our case, posts including news URLs pertaining to a news story) occur on a set of processes (in our case, Web communities).
A Hawkes model consists of $K$ point processes, each with a \emph{background rate} of events $\lambda_{0,ik}$.
Note that the events considered for Hawkes processes are a set of posts made on Web communities, and do not have to be confused with the event IDs from the GDELT dataset, that we used to identify the news stories.
For us, the point processes will be the timeseries $\{u_{ik}|k=1,\dots,K\}$ for a given story $i$.
The background rate is the expected rate at which events referring to a story will occur on a process {\em without} influence from the processes modeled or previous events; this captures stories mentioned for the first time, or those seen on a process we do not model and then occur on a process we do model.
An event on one process can cause an \emph{impulse response} on other processes, which increases the probability of an event occurring above the processes' background rates.
The shape of the impulse response determines how the probability of these events occurring is distributed over time.
Hawkes Processes are used for various tasks like modeling the influence of specific accounts~\cite{alvari2019hawkes,zannettou2019let,zannettou2019disinformation}, quantifying the influence between Web communities~\cite{zhou2013learning,zannettou2017web,zannettou2018origins}, and modeling information diffusion~\cite{soni2019detecting,kong2019linking,guo2015bayesian,lukasik2016hawkes}.
Here, we use them to quantify the influence between multiple Web communities with respect to the dissemination of news.

\descr{Model fitting.} We assume a Hawkes model that is fully connected: each process can influence all the others, as well as itself, which describes behavior where a user on a Web community sees a news story and re-posts it on the same platform.
Fitting a Hawkes model to a series of events on a set of processes provides us with the values for the background rates for each process, along with the probability of an event on one process causing events on other processes.
Background rates also let us account for the probability of an event caused by external sources of information---i.e., a Web community that we do not model.
Thus, while we only model the influence for a limited number of Web communities, the resulting probabilities are affirmatively attributable to each of them as the influence of the greater Web is captured by the background rates.
This also helps addressing limitations of our datasets. In particular, the tweets that are not included in the Twitter $1\%$ Streaming API is absorbed into the background rate of each community, avoiding to erroneously attribute an event to a different community.
To fit a Hawkes model for each news story, we use the approach described in~\cite{linderman2014,lindermanArxiv}, which uses Gibbs sampling to infer the parameters of the model from the data, including the background rates and the shape of the impulse responses between the processes.

\descr{Influence.} Overall, this enables us to capture the interplay between the posting of news stories across multiple Web communities and quantify the influence that each Web community has on each other.
More precisely, we use two different metrics: 1)~\textbf{\em raw influence}, which can be interpreted as the percentage of news story appearances that are created on a \emph{destination} Web community in response to previously occurring appearances on a \emph{source} one; and 2)~\textbf{\em normalized influence} (or efficiency), which normalizes the raw influence with respect to the number of news story appearances on the source Web community, hence denoting how efficient a community is in spreading news stories to other Web communities.

\begin{figure}[t!]
\centering
\includegraphics[width=0.75\columnwidth]{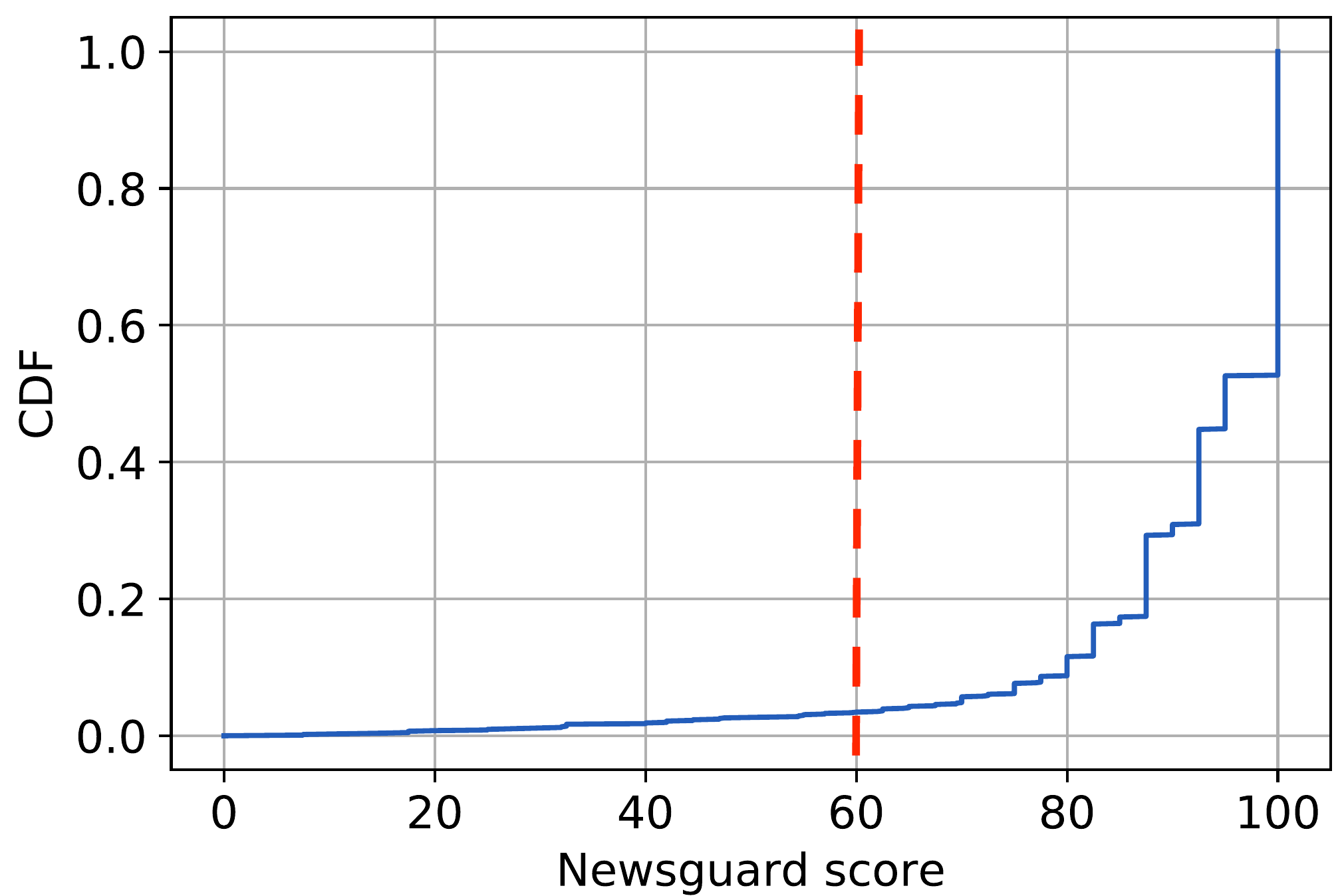}
  \caption{CDF of the NewsGuard scores of the news sources.}
\label{fig:cdf_newsguard_scores}
\end{figure}

\section{General Characterization}
\subsection{News Sources} %
Using the NewsGuard scores, we find that out of the 1,073 news sources in our dataset, 1,036 (96.6\%) are labeled as trustworthy (e.g., the New York Times, the Washington Post) and only 37 (3.4\%) untrustworthy (e.g., Infowars, Breitbart), i.e., they have a score of less than 60/100.
\figurename~\ref{fig:cdf_newsguard_scores} plots the CDF of the trustworthiness scores: most sources obtain relatively high scores, with 69\% of outlets scoring above 90, and almost half (47\%) receiving 100.
However, out of the 14M URLs in our dataset,
996K are to untrustworthy and 13M trustworthy news sources.
That is, over 7\% of posted URLs are from untrustworthy news even though these only account for 3.4\% of the sources.
Recall that the threshold of 60 is pre-defined by NewsGuard and is used as a guideline by their experts to rank news organizations.
The threshold value is an important factor when designers assign the points for each criterion.
For instance, even if a news outlet meets all transparency criteria (e.g., clearly lists funders) but fails all credibility criteria (e.g., repeatedly published false content and does not properly publish retractions), it would still receive a NewsGuard score of 25 and be therefore considered untrustworthy.
Any change of threshold need to reevaluate the points for each criterion at the same time, which is out the scope of this paper.
For this reason, it would not make sense for us to select a different threshold in this study.

\subsection{Named Entities}
Next, we describe the named entities extracted as per the methodology described in Section~\ref{sec:natural}.
Note that although GDELT does offer extracted entities in their metadata, %
we find that their labeling is not suitable for our purposes.
More specifically, GDELT relies on two databases of public figures which were last updated in 2010~\cite{tabari2014dictionaries}. %
So, for example, ``Trump'' does not appear in any of the entity metadata.
Instead, we use TrueCaseAnnotator, SpaCy, and \emph{en\_core\_web\_lg}.
Next, we describe the named entities extracted from the news articles in our dataset, and then move to the one extracted from the posts on Web communities containing news URLs.

\begin{table}[t]
\centering
\small
\begin{tabular}{lrlr}
\toprule
\multicolumn{2}{c}{\textbf{Trustworthy}} & \multicolumn{2}{c}{\textbf{Untrustworthy}}\\ \midrule
            \textbf{Entity} &   \hspace{-1.5cm}\textbf{(\%, out of 12M)} &             \textbf{Entity} &  \hspace{-1.5cm}\textbf{(\%, out of 920K)} \\
\midrule
             Trump &       \multicolumn{1}{r|}{17.94\%} &              Trump &       27.52\% \\
              U.S. &       \multicolumn{1}{r|}{15.53\%} &                 US &       18.74\% \\
          American &       \multicolumn{1}{r|}{11.15\%} &       Donald Trump &       18.30\% \\
      Donald Trump &       \multicolumn{1}{r|}{11.08\%} &           American &       15.54\% \\
  United States &       \multicolumn{1}{r|}{10.13\%} &               U.S. &       13.76\% \\
        Republican &       \multicolumn{1}{r|}{ 9.03\%} &             Russia &       13.39\% \\
        Washington &       \multicolumn{1}{r|}{ 8.44\%} &  United States &       13.14\% \\
           America &       \multicolumn{1}{r|}{ 7.24\%} &            America &       11.54\% \\
          New York &       \multicolumn{1}{r|}{ 6.87\%} &            Russian &       11.14\% \\
         Americans &       \multicolumn{1}{r|}{ 6.75\%} &              Obama &       10.20\% \\
           Reuters &       \multicolumn{1}{r|}{ 6.52\%} &          Americans &        9.77\% \\
          Congress &       \multicolumn{1}{r|}{ 6.18\%} &         Republican &        9.50\% \\
       Republicans &       \multicolumn{1}{r|}{ 6.11\%} &         Washington &        9.26\% \\
             Obama &       \multicolumn{1}{r|}{ 5.98\%} &          Democrats &        8.81\% \\
                US &       \multicolumn{1}{r|}{ 5.96\%} &           Facebook &        8.44\% \\
        Democratic &       \multicolumn{1}{r|}{ 5.90\%} &    Hillary Clinton &        7.48\% \\
         Democrats &       \multicolumn{1}{r|}{ 5.78\%} &           Congress &        7.46\% \\
            Russia &       \multicolumn{1}{r|}{ 5.73\%} &        Republicans &        6.95\% \\
          Facebook &       \multicolumn{1}{r|}{ 5.71\%} &              Syria &        6.73\% \\
           Twitter &       \multicolumn{1}{r|}{ 5.53\%} &            Twitter &        6.66\% \\
   White House &       \multicolumn{1}{r|}{ 5.42\%} &            Clinton &        6.47\% \\
\bottomrule
\end{tabular}
\caption{Top 20 named entities extracted from news articles originating from trustworthy and untrustworthy sources.}
\label{tbl:top_entities_mainstream_alternative}
\end{table}

\begin{figure*}[t]
\center
\subfigure[Twitter]{\includegraphics[width=0.495\textwidth]{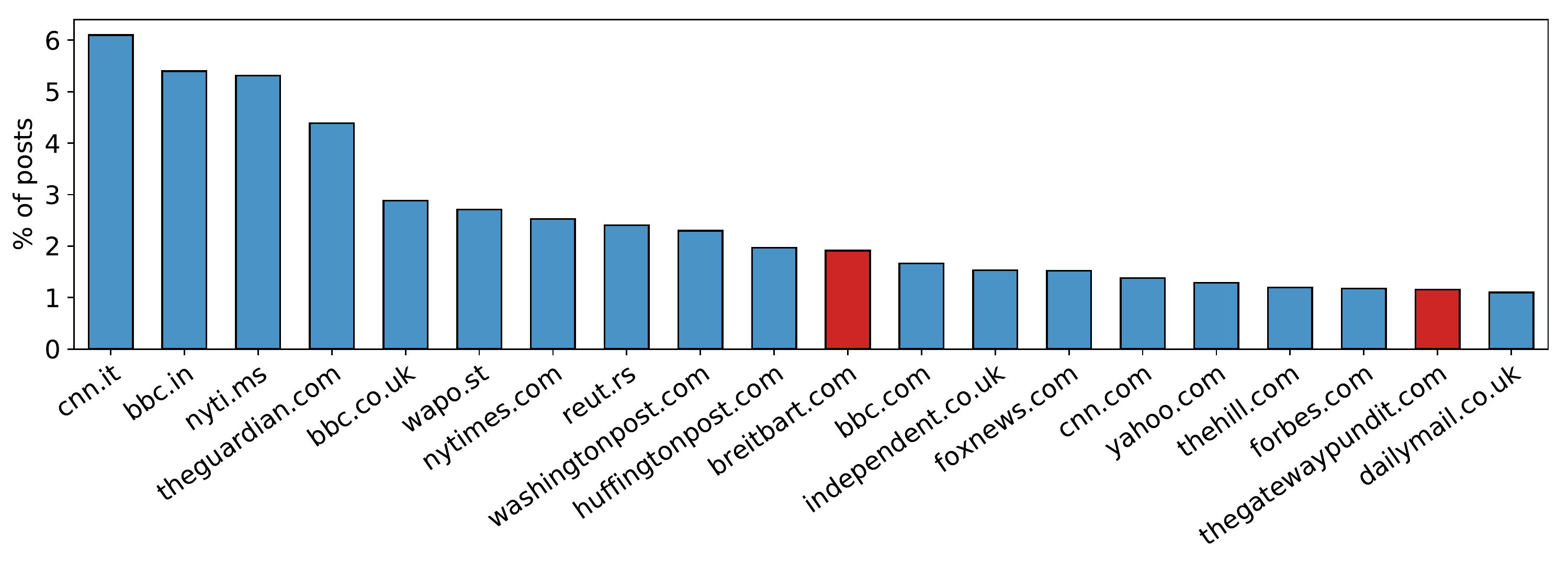}\label{bc_top_domains_twitter}}
\subfigure[Reddit]{\includegraphics[width=0.495\textwidth]{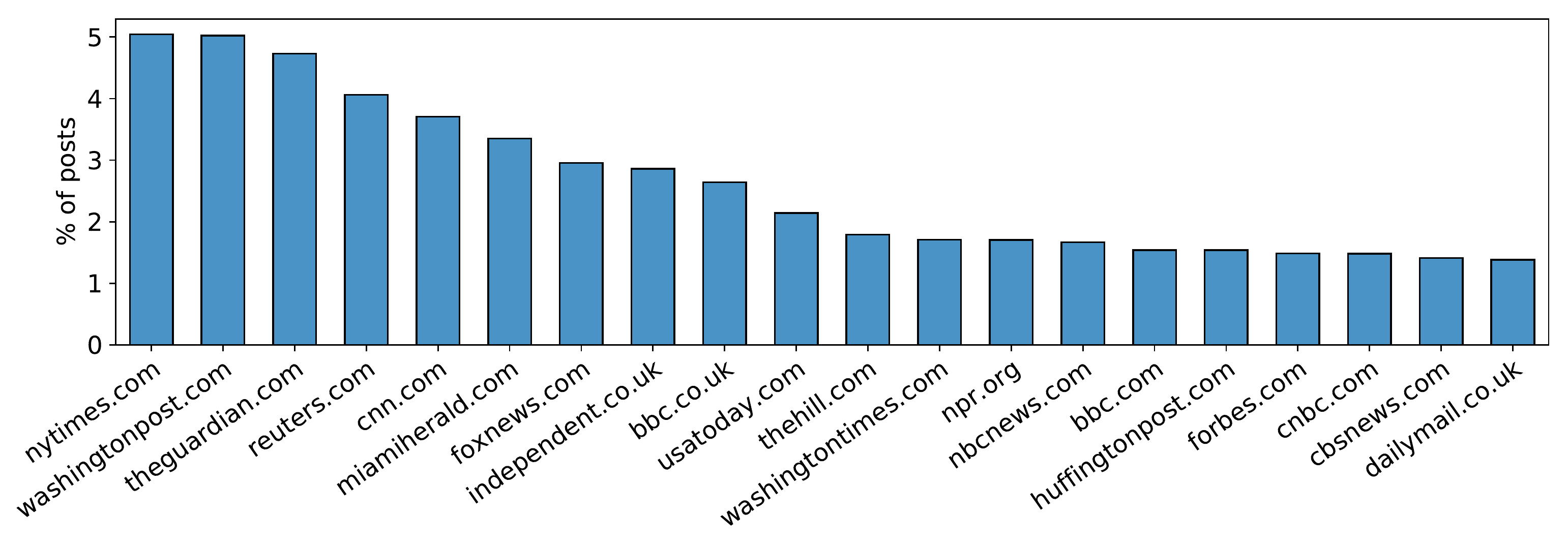}\label{bc_top_domains_reddit}}
\subfigure[4chan]{\includegraphics[width=0.495\textwidth]{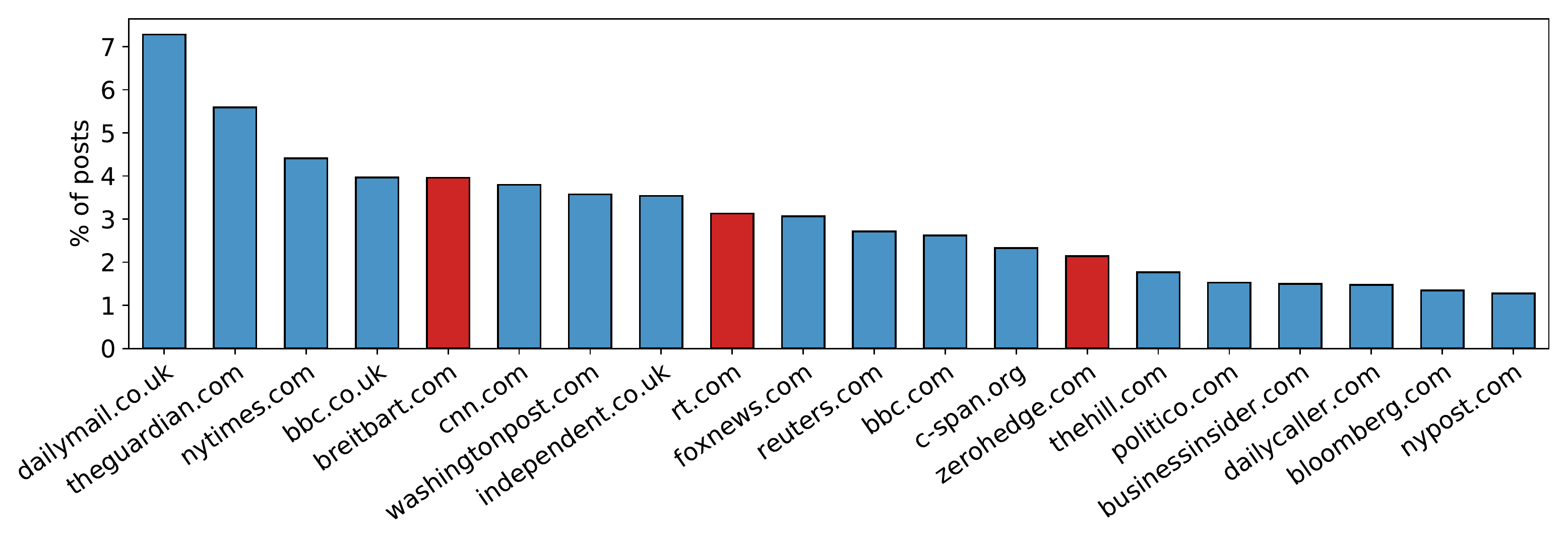}\label{bc_top_domains_4chan}}
\subfigure[/r/The\_Donald]{\includegraphics[width=0.495\textwidth]{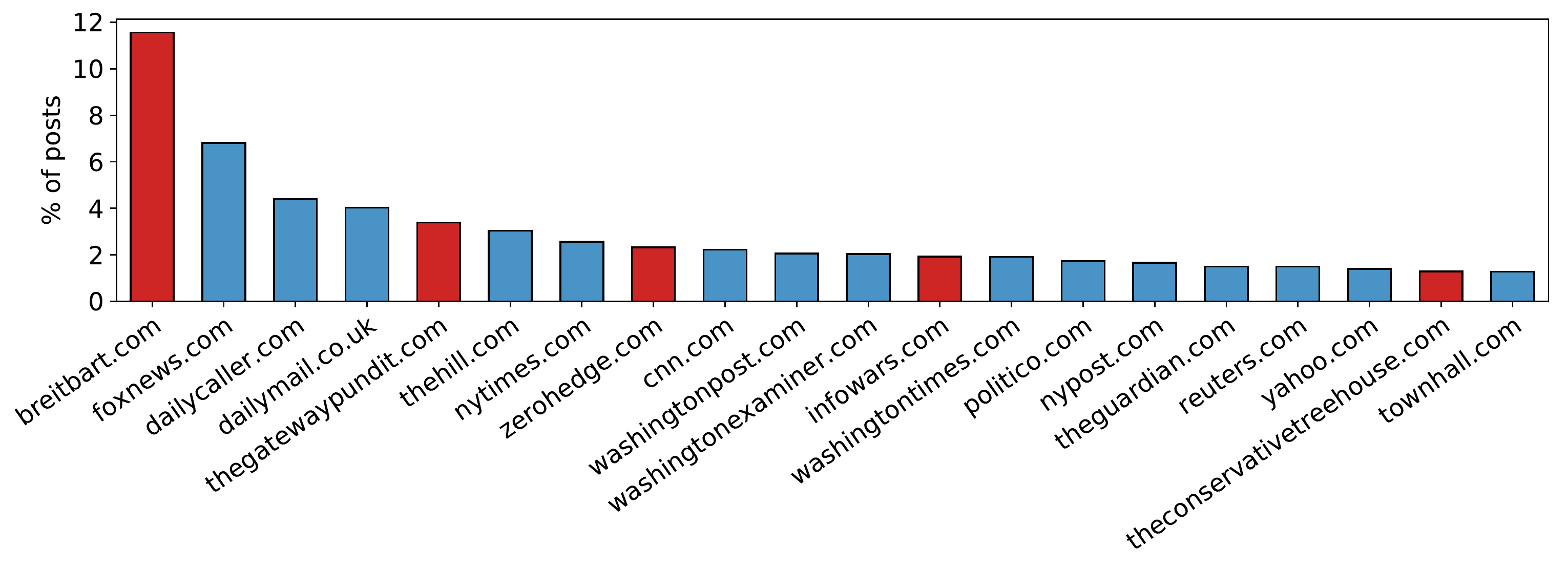}\label{bc_top_domains_td}}
\subfigure[Gab]{\includegraphics[width=0.495\textwidth]{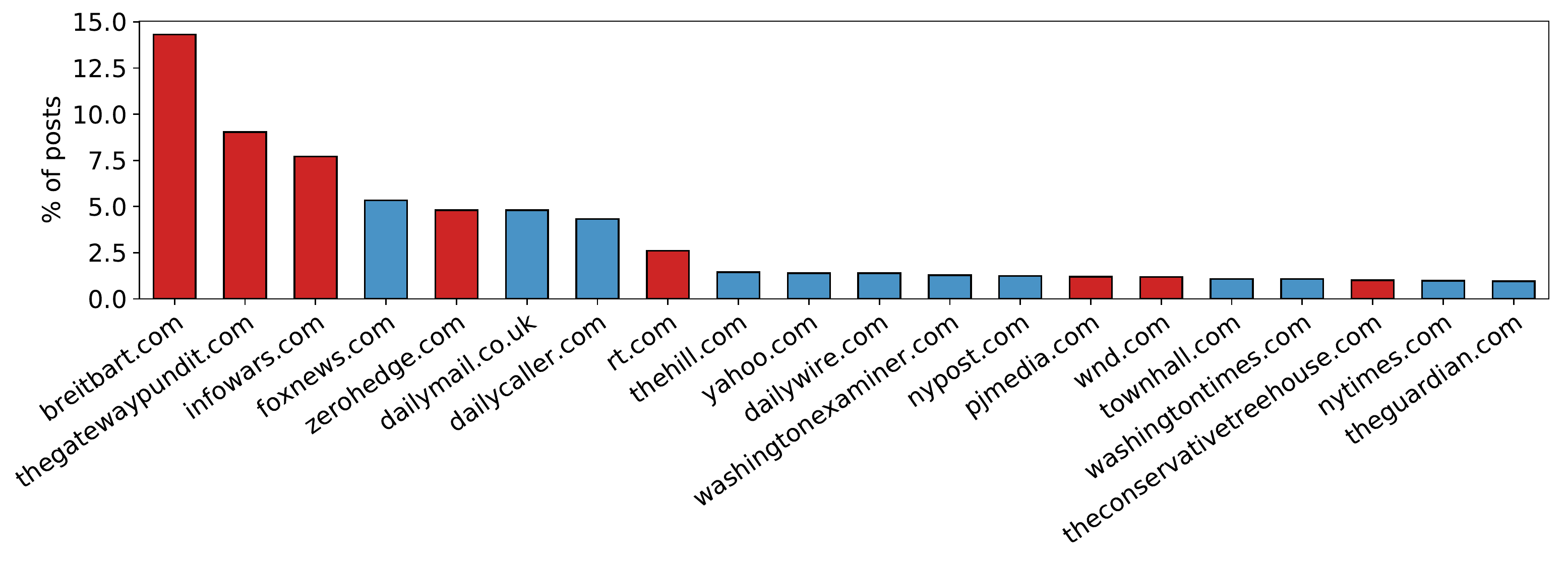}\label{bc_top_domains_gab}}
\caption{Top 20 domains according to their popularity. %
Blue and red bars denote, respectively, trustworthy and untrustworthy %
news sources.}
\label{bc_top_domains}
\end{figure*}

\descr{Articles.}
Using the article text and our pipeline's NLP component (Section~\ref{sec:natural}), we extract the named entities for each article in our dataset.
Table~\ref{tbl:top_entities_mainstream_alternative} reports the top 20 named entities extracted from both trustworthy and untrustworthy news articles.
We observe that the most popular entities referenced in both trustworthy and untrustworthy news are related to US politics.
For example, Donald Trump is referenced in 18\% and 27\% of the trustworthy and untrustworthy news articles, respectively.
We also find that both trustworthy and untrustworthy news articles mention {\em several} entities; about 90\% of them reference between 2 and 100 (the corresponding CDF plot is not included due to space limitations).
Finally, we note some differences between the top entities of trustworthy and untrustworthy news articles: for instance, ``Hillary Clinton'' appears in the latter but not in the former.

\begin{table*}[t!]
\centering
\vspace{0.5cm}
\footnotesize
\begin{tabular}{lrlrlrlrlr}
\toprule
\textbf{Twitter} & \multicolumn{1}{l}{\hspace{-0.75cm}\textbf{(\%, out of 7.8M)}} & \textbf{Reddit} & \multicolumn{1}{l}{\hspace{-0.75cm}\textbf{(\%, out of 24M)}} & \textbf{/r/The\_Donald} & \multicolumn{1}{l}{\hspace{-0.3cm}\textbf{(\%, out of 712K)}} & \textbf{4chan} & \multicolumn{1}{l}{\hspace{-0.75cm}\textbf{(\%, out of 521K)}} & \textbf{Gab} & \multicolumn{1}{l}{\hspace{-0.75cm}\textbf{(\%, out of 4.6M)}} \\ \midrule
Trump                     & \multicolumn{1}{r|}{7.67\%}                  & US                       & \multicolumn{1}{r|}{11.04\%}                 & Trump                         & \multicolumn{1}{r|}{9.46\%}                  & Trump                   & \multicolumn{1}{r|}{13.28\%}                 & Trump                 & 2.94\%                                       \\
US                        & \multicolumn{1}{r|}{1.46\%}                  & Trump                    & \multicolumn{1}{r|}{9.03\%}                  & Clinton                       & \multicolumn{1}{r|}{8.70\%}                  & US                      & \multicolumn{1}{r|}{10.45\%}                 & Obama                 & 2.08\%                                       \\
U.S.                      & \multicolumn{1}{r|}{1.26\%}                  & Russia                   & \multicolumn{1}{r|}{8.28\%}                  & Obama                         & \multicolumn{1}{r|}{8.05\%}                  & UK                      & \multicolumn{1}{r|}{8.36\%}                  & US                    & 1.97\%                                       \\
Donald Trump              & \multicolumn{1}{r|}{1.19\%}                  & Russian                  & \multicolumn{1}{r|}{6.27\%}                  & Hillary                       & \multicolumn{1}{r|}{7.29\%}                  & Israel                  & \multicolumn{1}{r|}{8.13\%}                  & FBI                   & 1.85\%                                       \\
Russia                    & \multicolumn{1}{r|}{1.02\%}                  & U.S.                     & \multicolumn{1}{r|}{5.47\%}                  & US                            & \multicolumn{1}{r|}{6.66\%}                  & Russia                  & \multicolumn{1}{r|}{7.75\%}                  & Democrats             & 1.42\%                                       \\
Obama                     & \multicolumn{1}{r|}{1.02\%}                  & China                    & \multicolumn{1}{r|}{4.03\%}                  & CNN                           & \multicolumn{1}{r|}{6.50\%}                  & EU                      & \multicolumn{1}{r|}{6.62\%}                  & America               & 1.32\%                                       \\
Clinton                   & \multicolumn{1}{r|}{0.95\%}                  & Clinton                  & \multicolumn{1}{r|}{3.88\%}                  & FBI                           & \multicolumn{1}{r|}{5.09\%}                  & U.S.                    & \multicolumn{1}{r|}{5.76\%}                  & CNN                   & 1.30\%                                       \\
GOP                       & \multicolumn{1}{r|}{0.91\%}                  & Obama                    & \multicolumn{1}{r|}{3.47\%}                  & Russia                        & \multicolumn{1}{r|}{5.01\%}                  & Russian                 & \multicolumn{1}{r|}{5.60\%}                  & Russia                & 1.16\%                                       \\
UK                        & \multicolumn{1}{r|}{0.79\%}                  & FBI                      & \multicolumn{1}{r|}{3.44\%}                  & Muslim                        & \multicolumn{1}{r|}{4.62\%}                  & Donald J                & \multicolumn{1}{r|}{5.42\%}                  & U.S.                  & 1.12\%                                       \\
CNN                       & \multicolumn{1}{r|}{0.68\%}                  & CNN                      & \multicolumn{1}{r|}{3.30\%}                  & Muslims                       & \multicolumn{1}{r|}{4.55\%}                  & TRUMPTV                 & \multicolumn{1}{r|}{5.39\%}                  & Muslim                & 1.10\%                                       \\
Hillary Clinton           & \multicolumn{1}{r|}{0.66\%}                  & American                 & \multicolumn{1}{r|}{2.93\%}                  & Hillary Clinton               & \multicolumn{1}{r|}{4.54\%}                  & American                & \multicolumn{1}{r|}{5.19\%}                  & American              & 1.04\%                                       \\
China                     & \multicolumn{1}{r|}{0.64\%}                  & Democrats                & \multicolumn{1}{r|}{2.82\%}                  & U.S.                          & \multicolumn{1}{r|}{4.36\%}                  & Jews                    & \multicolumn{1}{r|}{5.16\%}                  & UK                    & 1.02\%                                       \\
America                   & \multicolumn{1}{r|}{0.63\%}                  & UK                       & \multicolumn{1}{r|}{2.72\%}                  & American                      & \multicolumn{1}{r|}{4.23\%}                  & Syria                   & \multicolumn{1}{r|}{5.04\%}                  & Russian               & 0.84\%                                       \\
Russian                   & \multicolumn{1}{r|}{0.60\%}                  & GOP                      & \multicolumn{1}{r|}{2.59\%}                  & America                       & \multicolumn{1}{r|}{4.16\%}                  & Jewish                  & \multicolumn{1}{r|}{4.89\%}                  & Hillary               & 0.75\%                                       \\
Hillary                   & \multicolumn{1}{r|}{0.57\%}                  & Republicans              & \multicolumn{1}{r|}{2.45\%}                  & Islam                         & \multicolumn{1}{r|}{3.61\%}                  & China                   & \multicolumn{1}{r|}{4.66\%}                  & Americans             & 0.73\%                                       \\
FBI                       & \multicolumn{1}{r|}{0.56\%}                  & White House              & \multicolumn{1}{r|}{2.37\%}                  & Soros                         & \multicolumn{1}{r|}{3.25\%}                  & Clinton                 & \multicolumn{1}{r|}{4.44\%}                  & Clinton               & 0.71\%                                       \\
Republicans               & \multicolumn{1}{r|}{0.52\%}                  & America                  & \multicolumn{1}{r|}{2.35\%}                  & Democrats                     & \multicolumn{1}{r|}{3.17\%}                  & Brexit                  & \multicolumn{1}{r|}{4.19\%}                  & EU                    & 0.70\%                                       \\
BBC News                  & \multicolumn{1}{r|}{0.51\%}                  & Putin                    & \multicolumn{1}{r|}{2.33\%}                  & Donald Trump                  & \multicolumn{1}{r|}{2.85\%}                  & Britain                 & \multicolumn{1}{r|}{3.78\%}                  & Google                & 0.70\%                                       \\
Democrats                 & \multicolumn{1}{r|}{0.51\%}                  & Syria                    & \multicolumn{1}{r|}{2.28\%}                  & Russian                       & \multicolumn{1}{r|}{2.81\%}                  & Obama                   & \multicolumn{1}{r|}{3.71\%}                  & Democrat              & 0.68\%                                       \\
Brexit                    & \multicolumn{1}{r|}{0.50\%}                                       & Washington Post          & \multicolumn{1}{r|}{2.23\%}                                       & Facebook                      & \multicolumn{1}{r|}{2.50\%}                                       & Israeli                 & \multicolumn{1}{r|}{3.70\%}                                       & POTUS                 & 0.68\%                                       \\ \bottomrule
\end{tabular}%
\caption{Top 20 entities in posts that contain URLs to news articles.}
\label{tbl:top_entities}
\vspace{1cm}
\end{table*}

\descr{Web Communities.}
\label{sec:general_communities}
We also study the named entities that appear in posts including news URLs.
Note that these entities are \emph{not} related to the text of the news article pointed by the URL, but rather the comment it was posted with.
Table~\ref{tbl:top_entities} reports the top 20 named entities detected for each of the five Web communities.
Similar to the named entities detected on the news articles, most of the entities appearing in the posts are related to world events and politics, and in particular US politics.
For instance, one of the most popular entities is ``Trump,'' with 7.6\%, 9\%, 9.4\%, 13.2\%, and 2.9\%, for Twitter, Reddit, /r/The\_Donald, 4chan, and Gab, respectively.

There are also some interesting differences between top entities across the communities:
e.g., on 4chan, several entities are Jewish and Israel related
(Israel with 8\%, Jews 5\%, Jewish 4.8\%, and Israeli with 3.7\%), while, on /r/The\_Donald and Gab, we find Islam-related entities (``Muslim'' with 4.6\% on  /r/The\_Donald and 1.1\% on Gab).

\subsection{News Domains in Web Communities}
\label{Posting of News URLs in Web Communities}
Next, we study the popularity of the news sources on each Web community.
Overall, we find that 8.7\%, 5.0\%, 25.7\%, 12.6\%, and 48.7\% of the total occurrences point to untrustworthy URLs for Twitter, Reddit, /r/The\_Donald, 4chan, and Gab, respectively, while the rest point to trustworthy URLs.
In \figurename~\ref{bc_top_domains}, we report the top 20 news sources, in terms of their appearance, on each community.
Untrustworthy news sources are highlighted in red.

On Twitter (\figurename~\ref{bc_top_domains_twitter}) and Reddit (\figurename~\ref{bc_top_domains_reddit}), the most popular domains are mainstream, trustworthy news sources like The New York Times, The Washington Post, and CNN.
On /r/The\_Donald(\figurename~\ref{bc_top_domains_td}), the most popular news source is Breitbart, which was considered an untrustworthy news source from NewsGuard at the time of our experiments (57 score).
We also find other untrustworthy news sources, e.g., the Gateway Pundit (20 score), Zerohedge (0 score), and Infowars (25 score), among the top 12 most popular news sources.
We observe this phenomenon to an even greater extent on Gab (\figurename~\ref{bc_top_domains_gab}): the three most popular news sources are untrustworthy, with Breitbart being included in almost 15\% of Gab posts with news URLs.
For 4chan, we find mostly trustworthy news sources in the top 20, with the exception of Breitbart, the Russian state-sponsored RT (32.5 score), and Zerohedge.
The most popular news source is actually the Daily Mail, which has a 64.5 NewsGuard score.

Overall, these figures %
show that /r/The\_Donald and Gab are particularly polarized communities that extensively share news from untrustworthy sources, while Reddit and Twitter, which are more mainstream, do so to a much lesser extent.
4chan seems to be somewhat in the middle of the two: we find a substantial number of URLs from both trustworthy and untrustworthy news sources, which is perhaps surprising considering that 4chan is one of the most ``extreme'' communities on the Web~\cite{hine2017kek}.%

\begin{figure}[t!]
\centering
\includegraphics[width=0.8\columnwidth]{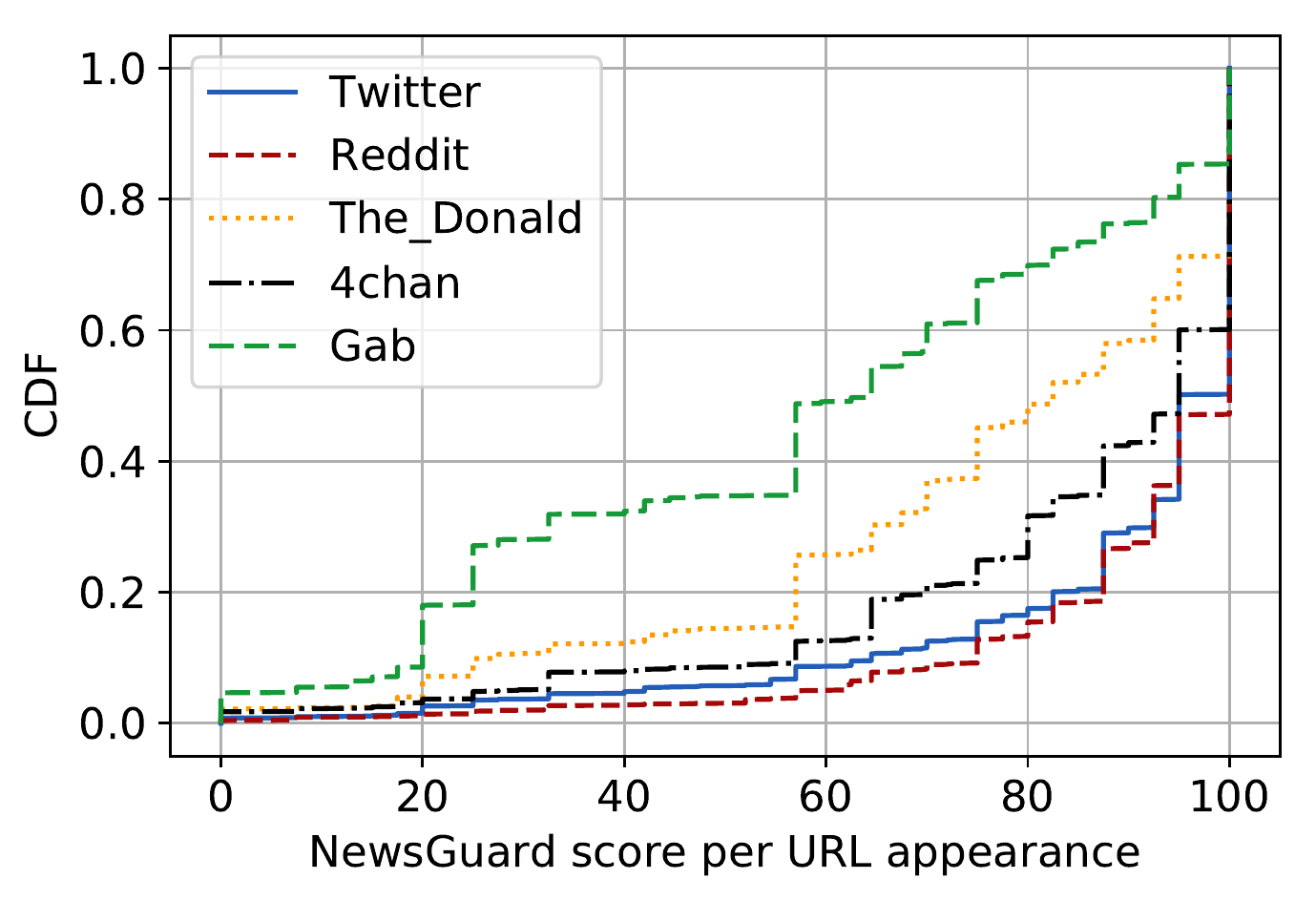}
\caption{CDF of the NewsGuard score for each news URL.} %
\label{fig:cdf_credibility_score_per_url}
\end{figure}

We also study trustworthiness at the granularity of specific URL appearances.
For each URL appearance, we extract the news source and assign its NewsGuard score.
In \figurename~\ref{fig:cdf_credibility_score_per_url}, we plot the resulting CDF.
Note that Gab shares substantially more URLs from less trustworthy sources (with a median score of 64.5), followed by /r/The\_Donald (median 82.5).
4chan shares substantially more URLs from %
trustworthy sources (with a median of 95) compared to Gab and /r/The\_Donald, and its median matches the one from Twitter (95).
Finally, Reddit users share predominantly URLs from trustworthy sources (with a median of 100).
To confirm these observations, we perform $\chi^2$ tests of independence on the proportion of trustworthy and untrustworthy news shared on each community.
The results allow us to reject the null hypothesis that there is no difference in the rate of trustworthy and untrustworthy shared by Web communities ($p<0.01$, with statistics values higher than 10,000 for all experiments).
\subsection{Main Takeaways}
Overall,  different Web communities discuss different types of news; e.g., 4chan focus more than others on Jewish and Israel related news, and /r/The\_Donald and Gab on news about Muslims.
Also, users on /r/The\_Donald and Gab prefer to cite untrustworthy news outlets to support their discussion.
We also perform a temporal analysis, aiming to capture the evolution of the trustworthiness scores as well as the interplay between trustworthy and untrustworthy news URLs on each platform.
Although we do not include the details in the paper due to space limitations, we find that the use of untrustworthy news outlets on Gab and /r/The\_Donald has been increasing over time, and on 4chan slightly decreasing, while remaining relatively stable on the other Web communities.

\section{Analyzing News Stories}
In this section, we set to understand how news stories, rather than single URLs, are discussed on different Web communities.
First, we describe the news stories identified, focusing on the differences across platforms. %
Then, we study the influence that different communities have on each other, using Hawkes Processes, and discuss a few interesting case studies.

\subsection{News Stories}
\label{sec:news stories and details}
Out of the 14M news URLs we extract from Twitter, Reddit, /r/The\_Donald, 4chan, and Gab, 3.2M of them appear in the GDELT dataset.
After the story identification, %
we extract 43,312 unique stories: 21,878  on  Twitter, 42,783 on Reddit, 4,943 on /r/The\_Donald, 5,007 on 4chan, and 9,929 on Gab.
These correspond to 109,153 unique URLs, 105,143 trustworthy and 4,010 untrustworthy.
Trustworthy URLs occur 130,235 times on Twitter, 469,058 on Reddit, 9,249 on /r/The\_Donald, 14,184 on 4chan, and 46,559 on Gab.
Untrustworthy URLs occur 2,759 times on Twitter, 9,221 on Reddit, 990 on /r/The\_Donald, 656 on 4chan, and 11,469 on Gab.
Recall that GDELT is focused on political stories with a national and international relevance.
As such, we expect news about local matters as well as sports or entertainment not to be included in this dataset.

\begin{figure}[t!]
\centering
\includegraphics[width=0.75\columnwidth]{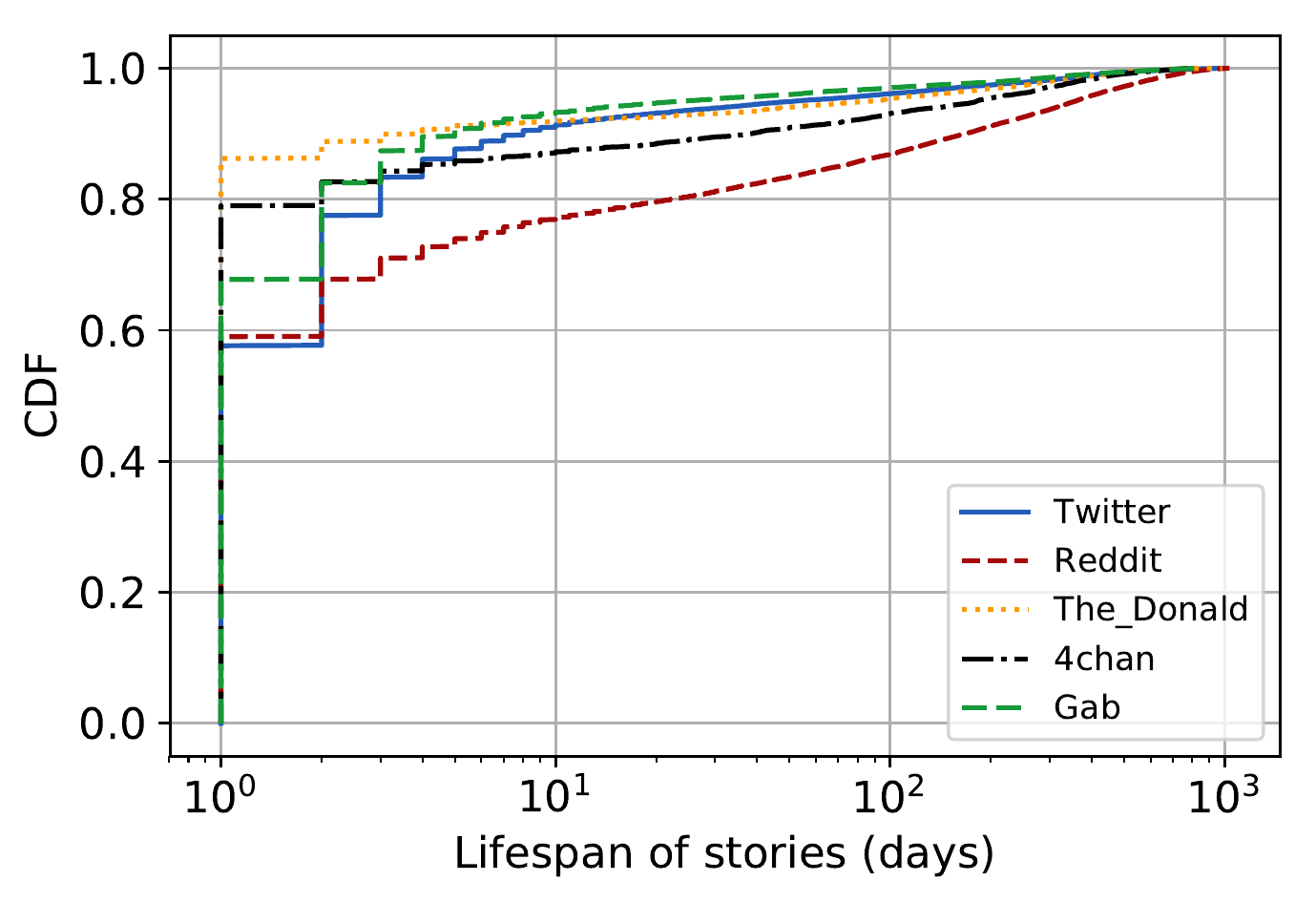}
  \caption{CDF of lifespan of stories on Web communities.}
\label{fig:Life span of different Web communities}
\end{figure}

Comparing this to the results in Section~\ref{Posting of News URLs in Web Communities}, we note that untrustworthy URLs appear in $2.1\%$ of all news story posts on Twitter (compared to $8.7\%$ for all news URL occurrences), while for Reddit this is $1.9\%$ (compared to $5.0\%$ overall), $4.4\%$ for /r/The\_Donald ($26\%$ overall), $4.4\%$ for 4chan ($13\%$ overall), and $21\%$ for Gab ($49\%$ overall).
This might indicate that although some communities prefer to use untrustworthy news URLs to support their discussion, when discussing political news stories they still prefer to quote trustworthy ones.
Our analysis in Section~\ref{sec:casestudies} suggests that this is sometimes done to give trustworthiness to a claim.

We then look at the {\em lifespan} of news stories on different Web communities, i.e., the time between the first and the last time a URL from a given story is posted on a platform.
Figure~\ref{fig:Life span of different Web communities} plots the CDF of the news story lifespans on the five Web communities.
The vast majority of stories on all platforms are short-lived, with a small number of notable exceptions; for instance, the story with the longest lifespan on Twitter (984 days) is about ``Saudi Arabia beheadings reaching the highest level in decades.''

Overall, Reddit users discuss news stories the longest, as almost one out of ten (8.8\%)  last for 200 days or more.
Interestingly, 4chan comes next, with 4.6\% of the stories being discussed on that platform for 100 days or more.
This is interesting, considering that posts on 4chan are ephemeral (i.e., they are deleted after a few days), with news content disappearing from the platform on a regular basis; hence, the fact that the 4chan community keeps discussing the same story for long periods of time indicates that new threads about it are constantly created.

\subsection{Influence Estimation} \label{sec:influence_results}
We now study the influence of Web communities w.r.t.~news stories.
As discussed in Section~\ref{sec:temporal}, we create a Hawkes model for each news story that appears at least 100 times on any platform, and more precisely 364 stories.
Note that each model consists of five processes, one per community.
Then, for each story, we fit a Hawkes model using Gibbs sampling.

Table~\ref{tbl:hawkes_events} reports the overall number of events (i.e., appearances of news stories) modeled with Hawkes Processes for each Web community.
Looking at the raw number of events, unsurprisingly, Reddit and Twitter are the communities with the most events in the selected 364 news stories.\footnote{Note that the start date of 4chan and Gab is behind the other Web communities (see Section~\ref{sec:methodology}). We find that $12.9\%$ of the analyzed stories occur before the collection period for these platforms. This accounts 8.3k out of 110k (7.5\%) events. Since only a minority of events are affected by this discrepancy, we consider the influence estimation experiments in this section to be an accurate reflection of real world trends.} %

\begin{table}[t]
\vspace{0.2cm}
\centering
\small
\begin{tabular}{r|r|r|r|r|r}
\toprule
\textbf{Twitter} & \textbf{Reddit} & \hspace*{-0.1cm}\textbf{/r/The\_Donald}\hspace*{-0.1cm} & \textbf{4chan}\hspace*{-0.1cm} &\hspace*{-0.1cm} \textbf{Gab}\hspace*{-0.1cm} & \textbf{Total} \\ \hline
24,987           & 65,610          & 1,926                & 3,163          & 11,800       & 107,486                                                          \\ \bottomrule
\end{tabular}%
\caption{Number of events %
modeled via Hawkes Processes.}
\label{tbl:hawkes_events}
\end{table}

\begin{figure*}[t!]
\center{}
\subfigure[]{\includegraphics[width=0.4\textwidth]{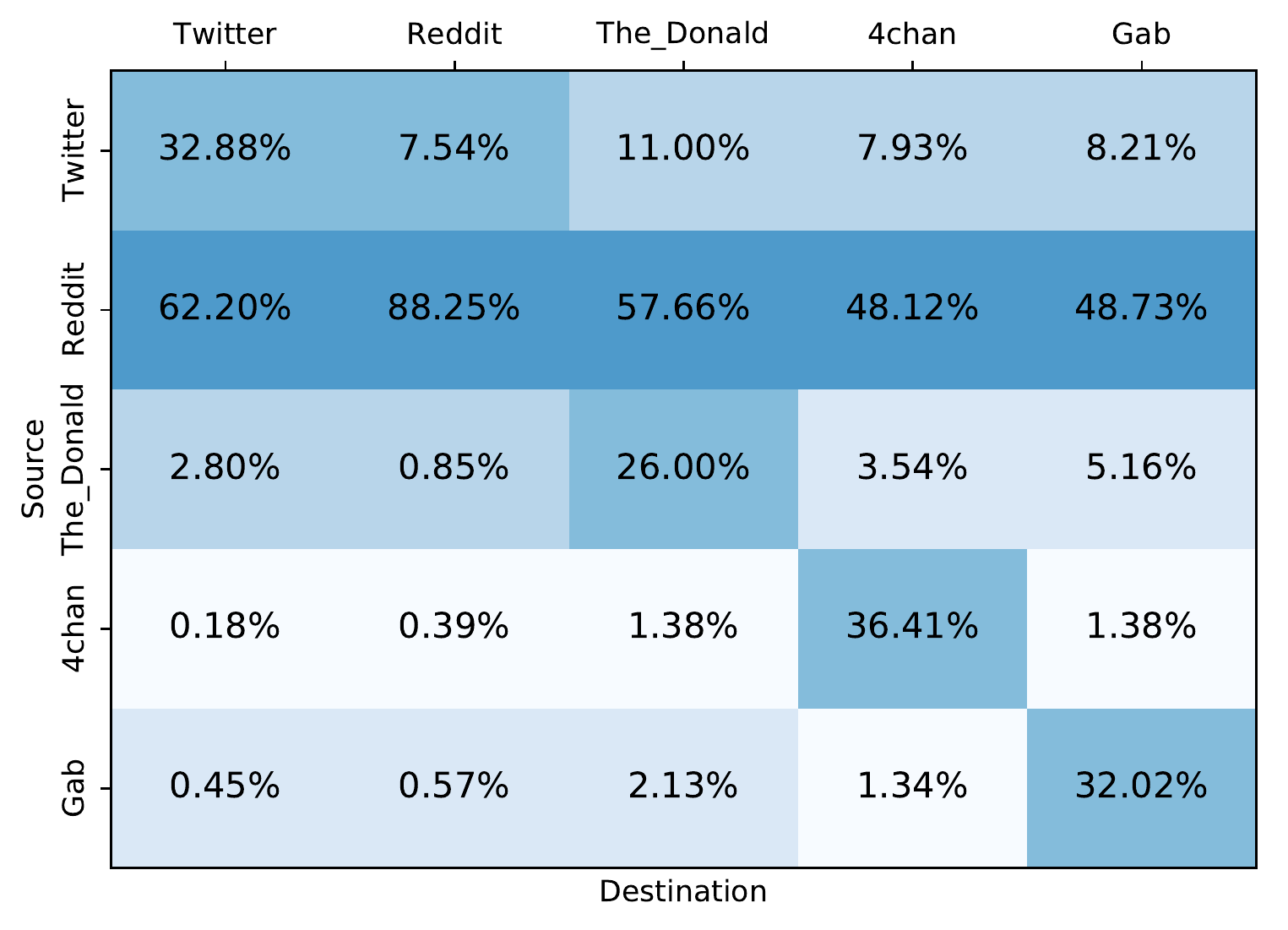}\label{subfig:hawkes_influence}}
\hspace{0.2cm}
\subfigure[]{\includegraphics[width=0.4\textwidth]{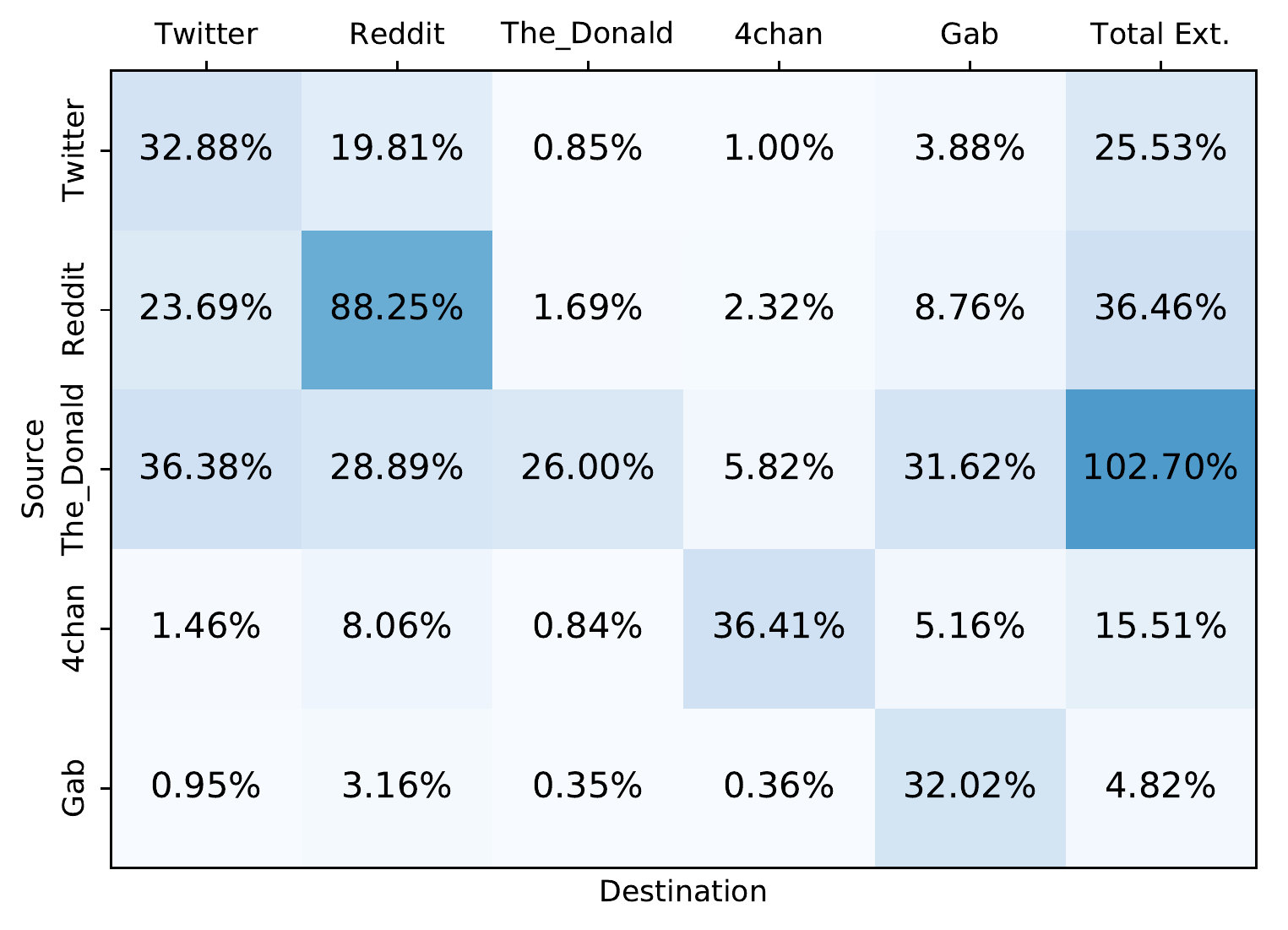}\label{subfig:hawkes_normalized}}
\caption{Influence estimation results: a) Raw influence between source and destination Web communities, which can be interpreted as the expected percentage of events created on the destination community because of previously occurring events on the source community; and b) Normalized influence (efficiency) of each Web community, which can be interpreted as the influence per news story appearance.} %
\label{fig:hawkes}
\end{figure*}

Fitting a Hawkes model provides us with the parameters for the background rates and impulse responses of each process, thus, we are able to quantify the influence that each Web community has on each other.
\figurename~\ref{fig:hawkes} shows our influence estimation results, which capture how influential and efficient Web communities are in spreading news stories.

Overall, we make several observations.
First, in terms of raw influence (see \figurename~\ref{subfig:hawkes_influence}), Twitter and Reddit are the most influential Web communities, mainly because of the large number of news story appearances that they produce.
Moreover, out of the three smaller communities (4chan, Gab, and /r/The\_Donald), /r/The\_Donald is the most influential Web community for news stories that appear on Twitter and Reddit.
This is particularly interesting since the overall number of news story appearances on /r/The\_Donald is substantially smaller compared to the ones on 4chan and Gab (see Table~\ref{tbl:hawkes_events}).
Finally, in terms of efficiency (see \figurename~\ref{subfig:hawkes_normalized}), /r/The\_Donald is by far the most efficient Web community in making news stories appear on other Web communities.

The last column in \figurename~\ref{subfig:hawkes_normalized} is the sum of normalized influence from the specific source community to the rest of the platforms and as such can be over $100\%$.
Generally, the bigger the percentage, the bigger the overall external influence from the source community to all the others; for instance, /r/The\_Donald has a high influence on all the other platforms, which adds up to over 100\%.

Note that statistical tests, e.g., to elicit confidence intervals  for the influence probabilities are hard to compute for Hawkes processes as, to the best of our knowledge, there is no statistical tool that is both meaningful and tractable.
More specifically, goodness of fit for Hawkes process exists but it has not been implemented or tested at scale and therefore we leave it as part of future work.

\subsection{Selected Case Studies}
\label{sec:casestudies}

Since our influence estimation is done on a per-story basis, we can identify stories for which each community is the most influential. %
This enables us to identify a few case studies we believe to be particularly interesting.
To do so, we calculate the overall external influence of each community and extract the top 20 externally influential stories (i.e., we do not include a community's self influence).
Given that a number of stories are influential on multiple Web communities, we obtain a set of 88 unique stories, which are the ones where either Twitter, Reddit, /r/The\_Donald, 4chan, or Gab is the most influential community on the other communities.
Below, we present some case studies from these 88 stories.

\descr{Presidential election.}
Out of these 88 news stories, 37 are related to the 2016 US Presidential Election.
Although the five Web communities under study all extensively discuss the election, we find that different communities push different narratives and topics onto other platforms.
For instance, Twitter influences the other platforms to discuss a story related to then-House Majority Leader Kevin McCarthy believing that Vladimir Putin was paying President Trump (e.g.,~\cite{twitter_election}),
while Reddit does so with respect to President Trump revealing classified information to the Russian Foreign Minister (e.g.,~\cite{reddit_election}).
/r/The\_Donald is influential in spreading a story suggesting that China hacked Hillary Clinton's email server (e.g.,~\cite{td_election}), while 4chan for a story about an Iowa woman arrested on suspicion of voting twice (e.g.,~\cite{4chan_election}).
For Gab we find that the community was influential for a story reporting that Trump won the vote in Wisconsin and Pennsylvania after a recount (e.g.,~\cite{gab_election}).

\descr{Immigration.} Over the past few years, refugee crises have been often covered in the news, as also confirmed by previous research~\cite{boudemagh2017news}. %
In our dataset, we find 11 news stories for which one Web community has influenced its spread on other platforms.
On Twitter, we find a story with articles about migrants stuck in US airports as a consequence of President Trump's immigration ban in 2017 (e.g.,~\cite{nyt_immigrants}).
For 4chan, we find misrepresentations of remarks made by the Mexican government at the NAFTA summit, interpreted as an acceptance to pay for the border wall (e.g.,~\cite{infowars_immigrants}), and a story alleging that being too lenient in accepting refugees made Sweden the ``rape capital of Europe'' (e.g.,~\cite{bbc_immigrants}).
Among the stories for which /r/The\_Donald has influence on other platforms, there is one claiming that the influx of immigrants is the cause of the rise in violent crime in Germany (e.g.,~\cite{bbc_immigrants2}).

\descr{Syrian conflict.} Another interesting set of (six) stories %
is centered around the Syrian conflict.
Again, the narratives differ greatly.
Twitter pushes a story on French officials confirming a chemical attack carried out by the Syrian government (e.g.~\cite{twitter_syria}), while /r/The\_Donald reports that the US dropped 26,141 bombs over Syria during the Obama Administration (e.g.,~\cite{td_syria}).
Gab has influence regarding a story on the US freezing funding to the White Helmets movement after false allegations of chemical attacks in Syria (e.g.,~\cite{gab_syria}).
This last example confirms the observations made by previous research on alternative narratives being assembled surrounding the White Helmets on social media~\cite{starbird2018ecosystem}.

\subsection{Main Takeaways}
In summary, we find that Twitter and Reddit are the most influential Web communities w.r.t.~discussing news stories.
However, /r/The\_Donald is the most efficient considered its small(er) size.
Our analysis also shows that different communities influence discussion on different stories.
In particular, while Twitter and Reddit do so for major events reported by mainstream news along ``neutral'' narratives, more polarized communities are influential in discussing stories with specific narratives, from celebrating or criticizing political figures~\cite{gab_election,td_syria}, to promoting anti-immigration rhetoric~\cite{bbc_immigrants2,bbc_immigrants} or distributing false news and conspiracy theories~\cite{td_election,infowars_immigrants,gab_syria}.

\section{Related Work}\label{sec:related}

\descr{News spread on social networks.}
Zhao et al.~\cite{zhao2011comparing} compare news topics on Twitter to traditional media, while others~\cite{lerman2010information, zannettou2017web} investigate how news from mainstream and alternative news sources spread on different Web communities.
Ratkiewicz et al.~\cite{ratkiewicz2011truthy, ratkiewicz2010detecting} present a service that aims to track the spread of political astroturfing on Twitter.
Vosoughi et al.~\cite{vosoughi2018spread} study the spread of true and false news on Twitter, finding that false news spread wider and faster than real news.
Shao et al~\cite{shao2018spread} analyze the role of bots in spreading false news on the Web, while Leskovec et al.~\cite{leskovec2009memetracking} characterize news articles by identifying textual memes in them.
Tan et al~\cite{tan2016lost} build a four-layer structure model to study how information spreads.
Zannettou et al.~\cite{zannettou2020news} study whether the appearance of news on Web communities like Reddit and 4chan affect the commenting activity of news articles with regards to hate speech.
A comprehensive survey on this line of work is available from~\cite{kumar2018false}.

Other researchers have also focused on specific events, and in particular on disinformation.
Wilson et al.~\cite{wilson2018assembling} analyze comments on Twitter around the Aleppo boy conspiracy theory, while Backfried et al.~\cite{backfried2016sentiment} investigate attitudes towards refugees in Europe on German media.
Starbird~\cite{starbird2017examining} studies how disinformation related to massive shooting events spreads on Twitter, and  Conover et al.~\cite{conover2011political} characterize Twitter users' political orientations based on tweets related to the 2010 U.S. Congress midterm elections.

Finally, Zannettou et al.~\cite{zannettou2017web} look at the occurrence of {\em single} URLs.
By contrast, we focus on organic discussion of news {\em stories} rather than single URLs, which lets us provide a comprehensive view of how news are discussed online.
We also cover significantly more news outlets (1073 vs 99), and more Web communities (Gab was not included in~\cite{zannettou2017web}).

Overall, our work is the first, to the best of our knowledge, to study how different Web communities discuss political news stories and how they influence each other in doing so.

\descr{News credibility.} Researchers have studied how to detect false news, focusing on the news story level, which is orthogonal to ours.
Typically, they formulate the problem as a classification task and use machine learning to solve it~\cite{wu2018tracing, shu2017fake, castillo2011information, wang2017liar}.
Another direction is to focus on the news outlet level; these two directions are often closely related.
As shown in~\cite{flintham2018falling}, the source of news plays a key role when people judge the authenticity of the story.
Several papers, e.g,~\cite{lazer2018science}, attempt to determine what are the untrustworthy news outlets.
Pennycook and Rand~\cite{pennycook2019fighting} assess the trustworthiness of a news outlet based on laymen's evaluations, %
showing that crowdsourced judgements are successful in assessing trustworthy news sources, although not as much as professional fact-checkers.

Gentzhow et al.~\cite{gentzkow2006media, gentzkow2015media} show that news outlets can report news in a biased way, which could mislead the readers; in fact,  Soraka et al.~\cite{soroka2019cross} demonstrate that people tend to be more ``attracted'' by negative news stories.
To reduce bias, Babaei et al.~\cite{babaei2018purple} propose a method to identify ``purple news,'' i.e., news that can be unanimously accepted by readers who have opposite political leanings.
Resnick et al.~\cite{resnick2018iffy} propose a metric, called the ``Iffy Quotient'', to evaluate the spread of untrustworthy news sources on Twitter and Facebook, also relying on NewsGuard.
In~\cite{grinberg2019fake, budak2019happened}, researchers collect false news outlets from various sources; Grinberg et al.~\cite{grinberg2019fake} find that, among 2016 Presidential Election voters on Twitter, only a small portion are exposed to and share news from untrustworthy news outlets.
Budak~\cite{budak2019happened} compares the prevalence of news from trustworthy and untrustworthy outlets, obtained from~\cite{allcott2019trends}, during the 2016 election campaign. Although news from trustworthy sources are shared the most, a growing number from untrustworthy outlets spreads over time.

Overall, our work differs from this line of work not only in terms of methodology, but also because, besides Twitter, we also study fringe, impactful communities like Gab and 4chan.
Moreover, using Hawkes processes, we are able to analyze the influence of fringe communities on mainstream ones, which may help to better understand the influence dynamics of false news sharing.%

\descr{Events recording databases.} Our work relies on %
NewsGuard~\cite{newsguardtech} to assess trustworthiness, GDELT~\cite{GDELT-EventCodebook} to find story events, etc.
Other sources in this context include the News API~\cite{newsapi.org} and Google news~\cite{news.google.com}.
Event Registry~\cite{leban2014event, rupnik2016news} is another service that aggregates news and provides insights to its users. %
Kwak and An~\cite{kwak2016two} compare GDELT to Event Registry, showing that the former contains a larger set of articles and is therefore more suitable for research.
Overall, GDELT has been extensively used by researchers to study topics related to refugees~\cite{boudemagh2017news}, protests~\cite{zhang2019casm}, the media landscape~\cite{rappaz2019dynamic}, objects in news pictures~\cite{kwak2016revealing}, and so on~\cite{gleditsch2014data, kwak2014first}.

\section{Conclusion}

This paper analyzed the sharing and the spreading of online news. %
We showed that different communities present fundamental differences; for instance, Gab and /r/The\_Donald ``prefer'' untrustworthy news sources (e.g., on Gab, 48.7\% of all news URLs are from untrustworthy sources, compared to the 8.7\% for Twitter).
We also found that smaller Web communities can appreciably influence the news discussion on larger ones, with /r/The\_Donald being very effective in pushing news stories on Twitter and the rest of Reddit.

Naturally, our work is not without limitations.
First of all, while we did our best to gather a view of online news discussion that was as comprehensive as possible,
our dataset of news websites only includes English news websites as identified by the Majestic list and NewsGuard.
Moreover, we focused our analysis to four social networks, i.e., leaving out others like Facebook, due to the difficulty of collecting data.
Finally, we relied on the GDELT dataset, which, as discussed, presented noise and crawling errors and on a named entity recognition model that is mostly trained on well-edited text like news articles. %
However, as discussed, we took several steps to mitigate these issues, by performing a sensitivity analysis that allowed us to build accurate communities of news articles to form the news stories that we analyzed.

Overall, our analysis builds on a novel, re-usable computational pipeline relying on tools from natural language processing, graph analysis, and statistics. %
As such, our approach to group related news together, track their discussion on multiple networks, and assess influence between Web communities in discussing them could serve as the foundation for a wealth of research not only in computer science, but also in journalism and political science.

As part of future work, we plan to expand our methodology to look at what language is used to discuss the same news story on different Web communities, and at whether or not using certain types of language (e.g., hate speech) has a particular influence in news discussion.

\small
\bibliographystyle{abbrv}
\bibliography{sigproc}  %

\end{document}